\documentclass[prd,twocolumn,aps,amsmath,amssymb,nofootinbib,
preprintnumbers,superscriptaddress]{revtex4}

\usepackage{amsmath,amssymb}
\usepackage{graphicx}
\IfFileExists{srcltx.sty}{\usepackage[active]{srcltx}}

{
{
{\newcommand{\Rb}{R_{p}\hskip-1.5mm\raisebox{-1.00mm}/}
\newcommand{\be}{\begin{equation}}
\newcommand{\ee}{\end{equation}}
\newcommand{\bea}{\begin{eqnarray}}
\newcommand{\eea}{\end{eqnarray}}

\begin{document}

\input epsf \renewcommand{\topfraction}{0.8}

\title{Inflation, baryogenesis and gravitino dark matter at ultra low reheat temperatures}
\author{Kazunori Kohri~$^{1}$, Anupam Mazumdar~$^{1,2}$, Narendra Sahu}

\affiliation{Physics Department, Lancaster University, LA1 4YB, UK,\\
$^{2}$~Niels Bohr Institute, Blegdamsvej-17, Copenhagen, DK-2100, Denmark}


\begin{abstract}
It is quite possible that the reheat temperature of the universe  is extremely low close to the scale of 
Big Bang nucleosynthesis, i.e. $T_{R}\sim 1-10$~MeV. At such low reheat temperatures generating matter anti-matter asymmetry and synthesizing dark matter particles are challenging issues which need to be addressed within a framework of beyond the Standard  Model physics. In this paper we point out that a  successful cosmology can emerge naturally provided the R-parity violating interactions are responsible for the excess in baryons over anti-baryons and at the same time they can explain the longevity of dark matter with the right abundance. 
\end{abstract}


\maketitle

Our universe might have gone through multiple phases of inflation, see for an example~\cite{Burgess}. It is 
paramount that the last phase of inflation must provide sufficient e-foldings of inflation to explain the large 
scale structure of the universe besides
providing the seed perturbations for the temperature anisotropy for the cosmic microwave background (CMB) 
radiation~\cite{Komatsu:2008hk}. It is also mandatory that a graceful exit of inflation must happen in such a way that 
the inflaton decay products must excite the Standard Model (SM) quarks and leptons~\cite{AEGM,AKM} required for 
the success of Big Bang Nucleosynthesis (BBN)~\cite{BBN-rev}. This can be achieved without any need of ad-hoc assumptions provided that the inflaton carries the Standard Model (SM) charges as in the case of~\cite{AEGM,AKM}.

The above construction is based on embedding  inflation within the {\it gauge invariant} flat directions of the 
Minimal Supersymmetric Standard Model (MSSM), for a review see~\cite{MSSM,Kusenko}. Since the inflaton
interactions are that of the SM, the inflaton directly decays into the SM quarks and leptons~\cite{AEGM,AKM}, and the lightest
supersymmetric particle (LSP)~\cite{ADM}. Note that the dark matter particles are created and matched with the current observations just from their thermal interactions. Moreover the reheat temperature is sufficiently high enough to create 
baryon asymmetry before the Standard electroweak transitions~\cite{MSSM,Kusenko}.

However there are also plethora of models of inflation which do not belong to the observable sector~\cite{Linde-book}. In such cases 
the inflaton belongs to the {\it hidden sector} whose mass and couplings a {\it priori} are not known to us. They can be either an {\it absolute  gauge singlet} or just a {\it SM gauge singlet}, whose couplings to the SM fields are generically considered to be the Planck suppressed. In this paper we call them {\it moduli}~\footnote{The word {\it moduli} is a misnomer here, as it 
may or may not bear its inkling to that of the string moduli arising from string compactifications~\cite{String}}.

The aim of the present paper is to provide a {\it minimal} and a {\it successful} cosmology prompting from a hidden 
sector physics which can explain baryogenesis and dark matter at ultra low reheat temperatures such as 
$T_R\sim 1-10$~MeV. Such a stiffest challenge can be posed by any inflationary model where the inflaton is a 
SM gauge singlet.

As we expect there would be many problems which we need to overcome. In any case 
the lowest mass of such a moduli field is always constrained by the success of BBN. 
The reheating temperature after the moduli decay  into the SM degrees of freedom 
is represented by
\begin{eqnarray}
    \label{eq:TR}
    T_{R} \sim \sqrt{\Gamma_{\phi} M_{\rm P}} = 10~{\rm MeV} 
    \left( \frac{m_{\phi}}{10^{2} {\rm TeV}} \right)^{3/2},
\end{eqnarray}
where we have used total decay width of the moduli $\Gamma_{\phi}
\sim m_{\phi}^{3}/M_{\rm P}^{2}$ with the reduced Planck mass $M_{\rm
P} \simeq 2.4 \times 10^{18}$ GeV. Since we request $T_{R}
\gtrsim $ 5 MeV, in order not to spoil the successes of BBN~\cite{Kawasaki:1999na}, 
we have a lower limit on the mass, $m_{\phi} \gtrsim 10^{5} {\rm GeV}$. Then, we get a 
relationship, $T_{R}/m_{\phi} \gtrsim 10^{-7}$.\\

\noindent
{\it \underline{Challenges for baryonic asymmetry}}:
If the moduli mass is heavier than $m_{\phi}\geq 10^{7.5}-10^{8}$ GeV
then it is possible to get a reheat temperature above $T_{R}\geq 100$~GeV.
At such reheat temperatures there are many ways to generate matter anti-matter
asymmetry such as electroweak baryogenesis, Affleck-Dine baryogenesis, thermal/non-thermal 
leptogenesis, baryogenesis via Q-ball evaporation, etc.~\cite{MSSM,Kusenko}.

The problem arises when the reheat temperature is of the order of $T_{R}\sim 1-10$~MeV. For such a 
low reheating many of the mechanisms for generating matter-anti-matter
asymmetry will not work. First of all the scale of thermalization and the hadronization ought to be
very close to each other~\cite{Jaikumar}, such that the universe could go through a successful 
phase of BBN. Secondly, one would have to directly create baryons and anti-baryons and 
the tiny asymmetry between them {\it simultaneously}. One can not resort to electroweak Sphaleron 
transitions whose rates are by now exponentially suppressed for $T\ll 100$~GeV.

In order to create baryon asymmetry we would require all the three well known Sakharov's conditions; 
(1) an out of equilibrium scenario, which can be obtained from the decay of the moduli, (2) baryon number 
violation and (3) sufficiently large CP asymmetry. The latter issues are challenging from model building point 
of view. Within SM, the  B and L are accidental global symmetries, therefore it is not clear that 
a {\it priori} B and L are conserved within MSSM. As we shall argue here, the only way one can obtain baryon
number violation if one breaks $R$-parity in the hadronic sector in such a way that it is well constrained by the 
present set of experiments.\\

\noindent
{\it \underline{The dark matter production}}:
In order to explain the large scale structures of the universe, we need to excite the dark matter. However exciting 
heavy thermal dark matter, such as a generic LSP mass of order $\sim 100$~GeV, is a challenging problem  
at such low reheat temperatures of order $T_R\sim 1-10$~MeV. Typically a thermal freeze out temperature will 
be proportional to the LSP mass, $m_{LSP}/20\sim 5-10$~GeV.  Therefore we would have to create LSP via 
non-thermal process  from the direct decay of the moduli. An important challenge arises when the $R$-parity 
is broken, then the LSP would potentially decay into quarks and leptons much before the structures can be 
formed in the universe. \\
 
 \noindent
{\it \underline{$R$-parity violation and baryogenesis}}:
Given the nature of the issues we are discussing here it is important to understand what are 
the current limits on $R$-parity violating interactions. Let us now consider a scenario where 
B and L are violated, then the MSSM superpotential allows the following well known gauge 
invariant terms:
\be
W_{\Rb}=\mu_i^{'} L_i H_u+\lambda_{ijk}L_i L_j \ell^c_k + \lambda^{'}_{ijk}L_i
Q_j d^c_k +\lambda^{''}_{ijk}u^c_i d^c_j d^c_k\,,
\label{rp-sup}
\ee
where $L_i=(\nu_i, \ell_i)$, $Q_i=(u_i,d_i)$, $H_u=(h_u^+, h_u^0)^T$, 
$H_d=(h_d^0, h_d^-)^T$, {\it etc} are $SU(2)_L$ doublets and $u^c_i$, 
$d^c_i$ are $SU(2)_L$ singlet quarks. In Eq.(\ref{rp-sup}), the first three terms violate 
lepton number by one unit ($\Delta L=1$), while the last term violates baryon number 
by one unit ($\Delta B=1$). For the stability of proton we assume that $\lambda_{ijk}=
\lambda'_{ijk}=0$. This can be accomplished if there exists any conservation of lepton 
number, which then forces $\mu'_i$ to be zero. Under this condition some of 
$\lambda^{''}_{ijk}$ couplings are considerably large. However, the nonobservation of 
certain phenomena give stringent constraints on these couplings. In particular, the 
electric dipole moment of neutron gives~\cite{Barbieri}
\be
{\rm Im}\left( \lambda''_{312} \lambda''_{332} \right) < 0.03 \left(\frac{0.01}{V_{td}} 
\right) \left( \frac{\tilde{M} }{{\rm TeV}} \right)^2 
\ee
Similarly the non-observation of $n-\bar{n}$ oscillation gives an upper bound on 
$\lambda^{''}_{11k}$ to be ~\cite{Barbieri}
\be
|\lambda''_{11k}| < \left( 10^{-6} - 10^{-5} \right) \frac{10^8 {\rm s} }{\tau_{\rm osc}} 
\left(\frac{\tilde{M}}{ \rm TeV} \right)^{5/2}
\ee 
Thus we see that $ \lambda''_{332}$ is hardly constrained and can be taken to be as large as
${\cal O}(1)$. We take this to our advantage in order to estimate the baryon asymmetry from the 
out of equilibrium decay of the moduli.

Let us consider that $\phi$ decays to MSSM degrees of freedom before BBN. Now due to the large 
branching ratio, the decay of $\phi$ mostly gives rise to gauge bosons and gauginos, although it 
decays to gravitino, fermion and sfermions with smaller branching ratios. Since there is a baryon 
number violation through the $R$-parity violating couplings $\lambda''_{ijk}$, the decay of 
moduli and its decay products, primarily gauginos, will produce a net baryon asymmetry. 

First of all note that within MSSM, the Planck scale suppressed decay of the moduli field, $\phi 
\rightarrow u_i u^c_j, d_i d^c_j$, does not give rise to a net CP violation up to one loop quantum 
correction. The CP asymmetry in the moduli decay arises only through the two loop quantum 
corrections which is suppressed in comparison to the CP asymmetry produced by the decay of 
gauginos. Therefore, in what follows we will discuss the baryon asymmetry from the decay of 
gaugino fields (gluino, Z-ino and photino), represented here as $\tilde{g}$. 

Let us  assume that the gauginos are heavier than the quarks and squarks. As a result their decay 
to a pair of quark and squark through one loop quantum correction gives rise to a net CP 
violation. The magnitude of CP violation in the decay:  $\tilde{g}\rightarrow t \tilde{t}^c$ 
can be estimated as~\cite{Cline}:
\bea
\epsilon &=& \frac{\Gamma \left( \tilde{g} \rightarrow t \tilde{t}^c \right) - 
\Gamma \left( \tilde{g} \rightarrow \bar{t} \tilde{t} \right) }
{\Gamma_{\tilde{g}}^{\rm tot}} \nonumber\\
& \approx & \frac{\lambda''_{323}}{16 \pi} \frac{ {\rm Im} \left( A_{323}^* m_{\tilde{g}} \right)} 
{|m_{\tilde{g}}|^2}
\label{cp_violation}
\eea
where $A_{323}$ is the trilinear SUSY breaking term and we also assume a maximal 
CP violation. As a result the decay of gauginos produce more squarks (antisqarks) than 
antisquarks (squarks). The baryon number violating ($\Delta B=1$) decay, induced by $\lambda''_{323}$ 
of squarks (antisquarks) to quarks (antiquarks) then gives rise to a net baryon asymmetry. Note that 
the decay of squarks (anti-squarks) are much faster than any other processes that would erase the 
produced baryon asymmetry. Hence the B-asymmetry can simply be given by:
\be
\eta_B\sim B_{\tilde{g}} \epsilon \frac{n_{\phi}}{s} \sim \frac{3}{4}B_{\tilde g}\epsilon 
\frac{T_{R}}{m_{\phi}}\,,
\label{B-asym}
\ee
where $B_{\tilde{g}}\sim 0.5$ is the branching ratio of the decay of $\phi$ to $\tilde{g} \tilde{g}$,
and in the above equation $s$ is the entropy density resulted through the decay of $\phi$. 
Let us consider a parameter space set by Eq.~(\ref{eq:TR}), where $T_R/m_{\phi}\sim 10^{-7}$ and 
$m_{\phi}\sim 10^{5}$~GeV. Therefore a reasonable CP violation of order $\epsilon\sim 0.01- 0.001$ 
could accommodate the desired baryon asymmetry of ${\cal O}(10^{-10})$ close to the temperature 
of $T\sim 10-1$~MeV.\\

\noindent
{\it \underline{Gravitino as a dark matter}}:
Let us now consider a possible dark matter candidate in our scenario. Due to violation of $R$-parity, as such
the LSP is not completely stable.  Therefore neutralino type standard dark matter scenario will not be an able  
candidate. Due to the large $R$-parity violating coupling, either arising from  $\lambda^{\prime\prime}_{332}$ 
or $\lambda^{\prime\prime}_{312}$, the neutralino will decay much before the age of the universe.
The only probable candidate for the dark matter would be the gravitino, whose lifetime will be further suppressed by the 
Planck suppressed interactions. Furthermore, if the gravitino is the LSP then the two body decay will be prohibited and the 
only viable channel will be the three body decay into the SM fermions, which will also include the $R$-parity violating 
coupling, i.e. $\lambda^{\prime\prime}_{323}$.

Let us now consider the gravitino abundance from the moduli decay:
\begin{eqnarray}
    \label{eq:abunGrav}
    Y_{3/2}  \sim B_{3/2} \frac{3T_R}{4m_{\phi}},
\end{eqnarray}
where $B_{3/2}$ is the branching ratio into  gravitino and would be
$B_{3/2} = 10^{-2} - 1$~\cite{GravFromModuli} with the mixing between
modulus and the supersymmetry-breaking filed. We have used an
approximation ${n_{\phi}}/{s} \sim
({3T_R}/{4m_{\phi}})$~~\footnote{The branching ratio of the gravitino
production from an absolute gauge singlet is little more contentious
than one would expect naively. The moduli decay rate could get a
helicity suppression which depends on the details of the SUSY breaking
hidden sector~\cite{GravFromModuli}. There are examples of hidden
sectors, where $B_{3/2}\sim 10^{-2}$, see the second and third references
in~\cite{GravFromModuli}.}.

Let us evaluate the  gravitino contribution to the density of the dark matter,
\begin{eqnarray}
    \label{eq:CDMcond}
    Y_{\rm 3/2} = 3 \times 10^{-10} 
    \left( \frac{m_{3/2}}{{\rm GeV}} \right)^{-1}
    \left( \frac{\Omega_{3/2}h^2}{0.11} \right),
\end{eqnarray}
where the density parameter of the present universe is reported by
WMAP 5-year to be $\Omega_{\rm CDM}h^2 \sim 0.11$~\cite{Komatsu:2008hk} 
with the normalized Hubble parameter $h$. Note that for a gravitino mass 
of order $1$~GeV we can explain the right dark matter abundance with 
$B_{3/2}\sim 10^{-2}$ and $T_R/m_{\phi}\sim 10^{-7}$.

In presence of $R$-parity violation it becomes important to ask whether the gravitino 
can live long enough to serve as a dark matter candidate or not.  One can estimate the
decay rate of the gravitino induced by the $R$-parity violation, which can be written as
\begin{eqnarray}
    \label{eq:Gamma32}
    \Gamma_{3/2} = 
    \frac{\lambda^{\prime\prime~2}_{323}}{192 \pi^{3}}
    \frac{m_{3/2}^{5}}{\tilde{M}^{2} M_{\rm P}^{2}},
\end{eqnarray}
where $\tilde{M}$ is the mass of the supersymmetric particles, i.e. sparticle,
which couples to gravitino and induces three-bodies decay. Eq.~(\ref{eq:Gamma32})
gives the lifetime of the gravitino,
\begin{eqnarray}
    \label{eq:gravlife}
    \tau_{3/2} \sim 2.3 \times 10^{22}{\rm sec} 
    \left( \frac{\lambda''_{323}}{0.1} \right)^{-2}
    \left( \frac{m_{3/2}}{{\rm GeV}} \right)^{-5}
    \left( \frac{\tilde{M}}{10^{3}{\rm GeV} } \right)^{2}.
\end{eqnarray}
Therefore, the lifetime of gravitino can be longer than the cosmic
age. However  there is an important point to note here. If the
gravitino mass is such that $m_{3/2}\leq 1$~GeV, then the gravitino is
absolutely stable as there is a kinematical suppression for a
gravitino to decay into the SM baryons. 

In addition, there is an attractive feature to note here that the gravitino 
production by the decay of other superparticles is also suppressed and negligible compared
with the direct 2-bodies decay of the moduli, except for the Next LSP
(NLSP) SUSY particles. They will be  produced by the moduli decay products, either they 
quickly decay into the NLSP directly or through some cascade decays without producing
gravitinos. Because of the $R$-parity violation, which induces the
3-bodies decay of NLSP into SM fermions, the lifetime of the Next LSP
(NLSP) can be much shorter than $10^{-2}$ sec, which evades the
strong BBN constraints~\cite{Kawasaki:2008qe}, with its decay width
$\Gamma \sim (\lambda''_{323})^{2} \alpha_{i}^{2} m_{\rm
NLSP}^{3}/\tilde{M}^{2}$ where $\alpha_{i}$ is the fine-structure
constant of the gauge coupling and $m_{\rm NLSP}$ is the NLSP mass. This
decay width is much larger than that into gravitino from NLSP, which
is suppressed by the Planck mass squared. Thus, the production mode of
the gravitino is dominated by the  decay of moduli into a pair of
gravitinos.

For a non-thermal creation of dark matter it is important to check the free
streaming length. The gravitinos can have a large velocity at the
radiation matter equality. However for the parameters we are
interested in the free streaming length comes out to be: $\lambda_{\rm
FS}\sim 0.1~{\rm Mpc}~\log_{e}(2L_{\rm max})~(m_{3/2}/1{\rm
GeV})^{-1}(m_{\phi}/10^{5}{\rm GeV})^{-1/2}$ with $L_{\rm max}={\cal
O}(10^{2})(m_{3/2}/1{\rm GeV})(m_{\phi}/10^{5}{\rm
GeV})^{1/2}$~\cite{NTWDM}. For $m_{3/2}\sim 1$~GeV, and $m_{\phi}\sim
10^{5}{\rm GeV}$, we obtain $\lambda_{\rm FS}\sim {\cal O}(0.1)-1$~Mpc. Such a free 
streaming length is marginal from the point of
view of growth in the dark matter fluctuations. The suppression of the
density contrast below the free streaming length results in erasing
small structures, which can be tested by comparison between  detailed
N-body simulations and observations of Lyman-$\alpha$ clouds, or
future submilli-lensing observation of sub-halos~\cite{Hisano:2006cj}.\\

\noindent
{\it \underline{A model for a hidden sector low scale inflation}}:
So far we have not discussed the cosmic role of a $\phi$ field. In our case the moduli can act as an inflaton. 
One can envisage a simple low scale inflationary model where inflation occurs near {\it the point of  inflection}
with a mass $m_{\phi}\sim 10^{5}$~GeV and a potential:
\begin{equation}
V(\phi)\sim \frac{m_{\phi}^2}{2}\phi^2-\frac{A\kappa}{6\sqrt{3}}\phi^3+\frac{\kappa^2}{12}\phi^4\,,
\end{equation}
where $A\approx 4m_{\phi}$ and $\kappa\sim 10^{-6}$. Inflation can happen
near $\phi_0\sim \sqrt{3}m_{\phi}/\kappa$
with an Hubble expansion rate, $H_{inf}\sim (m_{\phi}^2/\kappa M_P)\sim
10^{-2}$~GeV.  The amplitude of the density perturbations will be given
by~\cite{AKM}:  $\delta_H \approx (1/5\pi)(H_{inf}^2/\dot\phi)\sim
(\kappa^2M_P/3m_{\phi}){\cal N}^2\sim 10^{-5}$, where the  number of
e-foldings is given by: ${\cal N}\sim 50$. One of the dynamical
properties of an inflection point  inflation is that the spectral tilt
can be matched in a desired observable range: $0.92< n_s< 1.0$ for the
above  parameters~\cite{AEGM,ADM1,Lyth}.\\

\noindent
{\it \underline{Discussions}}:
It is also possible to imagine $\phi$ to be a curvaton~\cite{Enqvist},
which dominates the universe while decaying. It would be desirable to 
have a curvaton belonging to the observable sector~\cite{Kasuya}, but this need
not be the case always.  The curvaton model still requires the inflationary potential to
dominate the energy density initially, so that the curvaton remains
light during inflation. An observed amplitude of perturbations can be
created from the decay of the curvaton with a mass of order
$10^{5}$~GeV. If the curvaton oscillations dominate then there will be
no distinguishable CMB signatures except the spectral tilt is
generically $n_s\sim 1$. Since all of radiation, baryon, and dark matter
have the same adiabatic perturbations, our model should not be suffered
from the constraint from isocurvature perturbation (see the discussion
in~\cite{Lemoine:2009yu}).

Furthermore, one can also imagine to obtain a low scale baryogenesis via 
Affleck-Dine mechanism in a $R$-parity violating scenario with a moduli coupling
to $u^{c}_{i}d^{c}_{j}d^{c}_{k}$~\cite{Murayama}. However in such cases the $\phi$ field cannot 
act as an  inflaton. One would require an inflaton sector, and there will be an 
additional source for baryon isocurvature fluctuations which is already constrained by the current
WMAP data~\cite{Komatsu:2008hk}.

To summarize, we have realized a successful early universe cosmology within a {\it hidden sector} inflaton paradigm
which gives rise to seed perturbations for the CMB, an observable range of tilt in the power spectrum, and  ultra low 
scale reheat temperatures of order $1-10$~MeV. The origin of baryogenesis and dark matter in our scenario are
now related to the $R$-parity violating interaction of the type: $\lambda_{323}^{\prime\prime}u^{c}_{3}d^{c}_{2}d^{c}_{3}$. 
The baryonic asymmetry is created from the decay products of a singlet inflaton and a viable dark matter candidate is the 
gravitino. Future experiments such as electric dipole moment of neutron, dark matter searches and the upcoming LHC 
will be able to constrain our scenario by providing better handle on $R$-parity violating interactions.

The authors are partly supported by "UNIVERSENET" (MRTN-CT-2006-035863) and by STFC Grant PP/D000394/1.


\end{document}